\documentclass[aps,prd,superscriptaddress,twocolumn,natbib,nofootinbib]{revtex4-1}
\usepackage{aas_macros}
\usepackage{amsmath,amssymb}
\usepackage{graphicx} 
\usepackage{rotating}
\usepackage[normalem]{ulem}
\usepackage[caption=false]{subfig}
\usepackage{bm}
\usepackage{tabularx}
\usepackage[usenames]{color}
\usepackage{xspace}
\usepackage[plainpages=false, colorlinks=true, anchorcolor=blue, linkcolor=blue, citecolor=blue, bookmarks=false]{hyperref}
\newcolumntype{b}{X}
\newcolumntype{s}{>{\hsize=.5\hsize}X}

\def\be{\begin{equation}}
\def\ee{\end{equation}}
\newcommand{\bb}{\begin{bmatrix}}
\newcommand{\eb}{\end{bmatrix}}

\def\bea{\begin{align}}
\def\eea{\end{align}}

\def\be{\begin{equation}}
\def\en{\end{equation}}
\def\bea{\begin{eqnarray}}
\def\ena{\end{eqnarray}}

\newcommand{\steve}[1]{}

\begin{document}

\title{All correlations must die: \\Assessing the significance of a stochastic gravitational-wave \\background in pulsar timing arrays}

\author{S.~R.~Taylor}
\email[]{Stephen.R.Taylor@jpl.nasa.gov}
\affiliation{Jet Propulsion Laboratory, California Institute of Technology, 4800 Oak Grove Drive, Pasadena, CA 91106, USA}

\author{L.~Lentati}
\affiliation{Astrophysics Group, Cavendish Laboratory, JJ Thomson Avenue,  Cambridge, CB3 0HE, UK}
\author{S.~Babak}
\affiliation{Max-Planck-Institut f{\"u}r Gravitationsphysik, Albert Einstein Institut, Am M{\"u}hlenberg 1, 14476 Golm, Germany}
\author{P.~Brem}
\affiliation{Max-Planck-Institut f{\"u}r Gravitationsphysik, Albert Einstein Institut, Am M{\"u}hlenberg 1, 14476 Golm, Germany}
\author{J.~R.~Gair}
\affiliation{School of Mathematics, University of Edinburgh, King's Buildings, Edinburgh EH9 3JZ, UK}
\author{A.~Sesana}
\affiliation{School of Physics and Astronomy, University of Birmingham, Edgbaston, Birmingham B15 2TT, United Kingdom}
\author{A.~Vecchio}
\affiliation{School of Physics and Astronomy, University of Birmingham, Edgbaston, Birmingham B15 2TT, United Kingdom}

\date{\today}
\begin{abstract}
We present two methods for determining the significance of a stochastic gravitational-wave (GW) background affecting a pulsar-timing array, where detection is based on evidence for quadrupolar spatial correlations between pulsars. Rather than constructing noise simulations, we eliminate the GWB spatial correlations in the true datasets to assess detection significance with all real data features intact. In our first method, we perform random phase shifts in the signal-model basis functions. This phase shifting eliminates signal phase coherence between pulsars, while keeping the statistical properties of the pulsar timing residuals intact. We then explore a method to null correlations between pulsars by using a ``scrambled" overlap-reduction function in the signal model for the array. This scrambled function is orthogonal to what we expect of a real GW background signal. We demonstrate the efficacy of these methods using Bayesian model selection on a set of simulated datasets that contain a stochastic GW signal, timing noise, undiagnosed glitches, and uncertainties in the Solar system ephemeris. Finally, we introduce an overarching formalism under which these two techniques are naturally linked. These methods are immediately applicable to all current pulsar-timing array datasets, and should become standard tools for future analyses.
\newline
\newline
\end{abstract}


\maketitle

\section{Introduction}

The existence of gravitational waves (GWs) was recently confirmed with the detection of a binary black-hole merger by LIGO, ushering in the era of observational GW astronomy \citep{aa+16}. This detection relied on (amongst other factors) precision engineering, extensive theoretical development, and detailed detector noise characterization. The latter is important because it tells us how confident we are that the detector output contains a signal rather than a spurious noise feature. Methods for this (such as ``time sliding") are well-developed in the ground-based and space-based interferometry literature, but until recently have been lacking for pulsar-timing arrays \citep[PTA,][]{fb90}. We explore such methods here.

PTA searches rely on the expected GW-induced correlation signature between pulsars to discriminate the GW signal against noise. These noise processes can be intrinsic to each pulsar, such as intrinsic spin-noise due to rotational irregularities \citep[e.g.][]{sc10}, or delays in the pulse arrival time due to propagation through the interstellar medium \citep[e.g.][]{kc+13}.  Other noise processes can be correlated across pulsars, such as uncertainties in the Solar system ephemeris \citep{chm+10} which can induce a dipole-like correlation signature, and errors in clock standards \citep{hcm+12} which can induce a monopolar correlation signature. An isotropic stochastic gravitational-wave background (GWB) induces spatial correlations in the PTA that have a quadrupolar signature, known as the \textit{Hellings and Downs} curve \citep{hd83}. This signature is only a function of the angular separation between pairs of pulsars in the array, although there are more general correlation signatures for anisotropic backgrounds \citep{msmv13,tg13,grtm14}, and GWBs composed of non-GR polarisations \citep{cs12,grt15}.

Upper limits on an isotropic stochastic GWB from the three main pulsar-timing arrays (PPTA \citep{s+15}; EPTA \citep{ltm+15}; NANOGrav \citep{arz+15}) are now reaching the sensitivities required to constrain models of backgrounds generated by a population of supermassive black hole binaries \citep[e.g.,][]{svc08,s13,mop14,rwsh15,konl15}.  Recent projections suggest that there is significant probability that a stochastic GWB will be detected within the next decade \citep{sejr13,rsg15,tv+15}.

Several detection statistics exist for a GWB signal in pulsar timing data. Frequentist methods such as the ``optimal statistic" \citep{ccs+15,abc+09} measure how likely it is (in terms of number of standard deviations from zero) that a cross-correlated signal is present in our data rather than a common uncorrelated signal. As it is currently formulated, this statistic assumes that all cross correlated power comes from the GWB, although generalization to multiple spatially correlated processes is possible. Indeed, the Yardley statistic \citep{ych+11} has recently been modified to simultaneously fit for the presence of multiple spatially correlated signals \citep{thk+15}. 

Bayesian inference instead makes use of the fully-marginalized likelihood (or ``evidence") to determine the probability of one model over another. This allows for a statistically robust comparison of models that includes contributions to the correlated signal from GWs and e.g.\ clock or Solar system ephemeris errors. We can also perform model comparison in more general scenarios, when the correlation between pulsars has been modeled using either a smooth functional form \citep{tgl13}, or pairwise for each pulsar pair \citep{l+13}. The shortcoming of this approach is the explicit dependence on the appropriateness of the models being used for the evidence comparison. Objective reality is not probed by Bayesian analysis, only our formulated models, which should be as close as possible to being realistic representations of the underlying physics. If we formulate a series of poor models to test on the data, then Bayesian model selection will select the least poor, but this does not mean it is the best possible model. 

In this paper we present two approaches for determining the significance of GW-induced correlations present in pulsar timing data. The first approach exploits correlations in phase, where random phase shifts are introduced between pulsars to destroy signal phase coherence, but which preserve the statistical properties of the individual pulsar datasets. The second approach exploits the expected spatial correlation signature of a GWB. By ``scrambling" the positions of pulsars on the sky (and therefore their angular separations) we produce a template of the correlation signature that is effectively orthogonal to the signature of any true signal in the real dataset. In the following we use Bayesian methods, but these techniques can be straightforwardly applied to real datasets with frequentist detection statistics. 

Previous work in \citet{cs15b} has investigated the requirements of a robust detection of GWs in PTAs, focusing on issues regarding the position scrambling approach. In \citet{thk+15} several methods were explored to mitigate the influence of spatially-correlated noise on GWB detection significance, including $(a)$ fitting a clock error signal out of all the pulsars before the GW search; $(b)$ fitting a monopolar clock error simultaneously with the quadrupolar GW signal; $(c)$ fitting a dipolar ephemeris error simultaneously with the quadrupolar GW signal; $(d)$ fitting an ephemeris error time-series out of all pulsars before the GW search. In the following, we show that Bayesian methods can avoid inference bias by simultaneously modeling all spatially-correlated processes, intrinsic pulsar noise, and timing model parameters. ``Phase-shifting" and ``sky-scrambling" are designed to operate on actual pulsar-timing datasets to provide a more conservative estimate of detection significance than methods relying on noise simulations, which can be biased to higher significance if undiagnosed noise features are present.

In Section \ref{sec:pta_likelihood} we introduce the Bayesian pulsar-timing likelihood. In Section \ref{sec:kill_corr} we discuss the phase and spatial correlations of pulsars due to the influence of a GWB, and introduce our two methods of phase shifting and sky scrambling. These methods are applied in Sec.\ \ref{sec:sims} to $(1)$ an idealized simulation; $(2)$ a more realistic simulation; $(3)$ a simulation for which our noise model is incomplete, where there are large glitches in each pulsar; and $(4)$ a simulation which contains additional noise processes that induce spatial correlations, i.e.\ a clock and ephemeris error. With the latter two, we show the superiority of our methods over repeated noise-only simulations. We summarize our findings in Sec.\ \ref{sec:conclusions}.
 
\section{A pulsar timing likelihood} 
\label{sec:pta_likelihood}

For any pulsar we can write the times of arrival (TOAs) for the pulses as a sum of both a deterministic and a stochastic component:
\begin{equation}
\mathbf{t}_{\mathrm{tot}} = \mathbf{t}_{\mathrm{det}} + \mathbf{t}_{\mathrm{sto}},
\end{equation}
where $\mathbf{t}_{\mathrm{tot}}$ is a vector of length $N_\mathrm{TOA}$ for a single pulsar, with $\mathbf{t}_{\mathrm{det}}$ and $\mathbf{t}_{\mathrm{sto}}$ the deterministic and stochastic contributions (modeled as Gaussian random processes) to the total respectively. An initial estimate, $\bm{\beta_{\mathrm{0}}}$, for the $m$ timing model parameters for each pulsar can be obtained through a standard weighted least-squares fit, or using Bayesian analysis routines \citep{lah+14}, both of which are included in the \textsc{Tempo2} \citep{hem06,ehm06} timing package.  This allows us to generate an initial set of timing residuals, which we denote $\mathbf{\delta t}=\mathbf{t}_{\mathrm{tot}}-\mathbf{t}_{\mathrm{det}}(\bm{\beta_{\mathrm{0}}})$.

We assume that the difference between this initial solution $\bm{\beta_{\mathrm{0}}}$, and the final solution $\bm{\beta_{\mathrm{f}}}$ obtained from a joint analysis that includes a GWB term will be small. Therefore a linear approximation of the timing model can be used such that any deviations from the initial guess of the timing model parameters are encapsulated using the vector $\bm{\epsilon}$ of length $m$, such that $
\epsilon_i = \beta_{\mathrm{f}i} - \beta_{\mathrm{0}i}$. These small timing model deviations influence the timing residuals via the term $\mathrm{\bf{M}}\bm{\epsilon}$, where $\mathrm{\bf{M}}$ is the $N_\mathrm{TOA}\times m$ timing model ``design matrix" describing the dependence of the residuals on the timing model parameters.

Furthermore, we include the influence of all low-frequency processes on the timing residuals (such as intrinsic spin-noise, a common red-noise process, and a GWB) via the term $\mathbf{F}\mathbf{a}$. The vector $\mathbf{a}$ of length $2N_\mathrm{freqs}$ describes the Fourier coefficients of any low-frequency process at a limited number of harmonics of the base sampling frequency $1/T$ (where $T$ is the observation timespan of a single pulsar, or the maximum coverage of the entire pulsar timing array), and $\mathbf{F}$ is the $N_\mathrm{TOA}\times 2N_\mathrm{freqs}$ ``Fourier design matrix" consisting of alternating columns of sines and cosines.

We can also explicitly include the influence of white-noise terms on the timing residuals, such as from TOA measurement uncertainties (which may be modified by additional system-dependent scaling parameters such as EFACs and EQUADs), correlated measurement uncertainties in simultaneous multi-frequency observations (ECORR), or pulse phase jitter \citep{sod+14}. However we implicitly marginalize over these effects in the following such that their influence is confined to the $N_\mathrm{TOA}\times N_\mathrm{TOA}$ white noise covariance matrix, $\mathrm{\bf{N}}$, for each pulsar.

The model-dependent timing residuals, $\mathbf{r}$, for each pulsar can thus be written in terms of the input residuals, $\mathbf{\delta t}$ as
\begin{equation}
\mathbf{r} = \mathbf{\delta t} - \mathrm{\bf{M}}\bm{\epsilon} - \mathrm{\bf{F}}\mathbf{a},
\end{equation}
with a likelihood given by
\begin{equation}
p(\mathbf{\delta t}|\bm{\epsilon},\mathbf{a},\boldsymbol{\eta}) = \frac{\exp\left(-\frac{1}{2}\mathbf{r}^T \bf{N}^{-1} \mathbf{r}\right)}{\sqrt{\mathrm{det}(2\pi \bf{N})}},
\end{equation}
where $\boldsymbol{\eta}$ encapsulates any parameters not already represented by $\bm\epsilon$ or $\mathbf{a}$.

We group all low-frequency and reduced rank signals into a common description, such that
\begin{equation}
\mathbf{r} = \mathbf{\delta t} - \mathrm{\bf{T}}\mathbf{b},
\end{equation}
where 
\begin{equation}
\mathrm{\bf{T}} = \begin{bmatrix} \mathrm{\bf{M}} & \mathrm{\bf{F}} \end{bmatrix} \;,\; \mathbf{b} = \begin{bmatrix} \bm{\epsilon} \\ \mathbf{a} \end{bmatrix}.
\end{equation}

We place a Gaussian prior on the coefficients, $\mathbf{b}$:
\begin{equation}
p(\mathbf{b}|\bm{\phi}) = \frac{\exp\left(-\frac{1}{2}\mathbf{b}^T \bf{B}^{-1} \mathbf{b}\right)}{\sqrt{\mathrm{det}(2\pi \bf{B})}},
\end{equation}
with,
\begin{equation}
\bf{B} = \begin{bmatrix} \boldsymbol{\infty} & 0 \\ 0 & \boldsymbol{\varphi} \end{bmatrix},
\end{equation}
such that the timing model portion of $\mathbf{b}$ has an infinite variance to approximate a uniform unconstrained prior on timing model parameter deviations, $\bm{\epsilon}$.

The low-frequency portion of $\mathbf{b}$ has a variance, $\varphi$, given by the spectrum of all low-frequency processes in the data. Since this may include a GWB we must naturally model spatial correlations in the data:
\begin{equation} \label{eq:phi_sum}
[\varphi]_{(ai),(bj)} = \Gamma_{ab}\rho_i\delta_{ij} + \kappa_{ai}\delta_{ab}\delta_{ij},
\end{equation} 
where $\kappa_{ai}$ is the intrinsic low-frequency (``spin-noise") spectrum of pulsar $a$ at the $i^\mathrm{th}$ sampling frequency; $\rho_{i}$ is the GWB spectrum at the $i^\mathrm{th}$ sampling frequency; and $\Gamma_{ab}$ is the overlap reduction function (ORF) between pulsars $a$ and $b$ describing the reduction in correlated power due to the spatial separation of the pulsars. For an isotropic stochastic GWB this $\Gamma_{ab}$ depends only on the separation between pulsars and is commonly known as the ``Hellings and Downs curve". We note that both $\kappa$ and $\rho$ can either be modelled with a functional form (such as a power-law or a smooth turnover) or as a free spectrum with a parameter per frequency. In the following we consider all low-frequency processes to be well described by power-law spectra at all sampling frequencies $\nu_i$, such that (taking $\kappa_{ai}$ as an example)
\begin{equation}
\kappa_{ai} = \frac{A_a^2}{12\pi^2}\frac{1}{T} \left(\frac{\nu_i}{1\mathrm{yr}^{-1}}\right)^{-\gamma_a} \mathrm{yr}^2. 
\end{equation}

For a GWB the exponent has a value of $\gamma=13/3$ for a circular GW-driven population of SMBHBs. All intrinsic red noise and GWB power-law spectral parameters are grouped into the parameter vector $\bm{\eta}$. We can trivially include other spatially correlated signals in the model by adding additional terms to  Eq.\ (\ref{eq:phi_sum}), e.g.\  a monopolar-correlated process to model clock errors, or a dipolar-correlated process as a (sub-optimal) model of ephemeris uncertainties. Ephemeris uncertainties can be modeled coherently \citep{chm+10,dhy+13,ltm+15} rather than through spatial-correlation analysis.

We can now write the joint probability density of the timing model and reduced rank signal parameters, $p(\mathbf{b}, \bm{\eta} |\mathbf{\delta t})$, as:
\begin{equation}
\label{Eq:Prob}
p(\mathbf{b}, \bm{\eta} | \mathbf{\delta t}) \propto p(\mathbf{\delta t} |  \mathbf{b}) \times p(\mathbf{b} | \bm{\eta} ) \times p(\bm{\eta}).
\end{equation}

Taking the logarithm of Eq.\ (\ref{Eq:Prob}) and extremizing gives the maximum likelihood vector of coefficients $\widehat{\mathbf{b}}$:
\begin{equation}
\label{Eq:bmax}
\widehat{\mathbf{b}} = \bm{\Sigma}^{-1}\mathbf{d},
\end{equation}
where $\mathbf{\Sigma} = (\mathbf{T}^T\mathbf{N}^{-1}\mathbf{T} + \bm{\varphi}^{-1})$ and $\mathbf{d} = \mathbf{T}^T\mathbf{N}^{-1}\mathbf{\delta t}$.

We can also analytically marginalize Eq.\ (\ref{Eq:Prob}) over the coefficients $\mathbf{b}$, giving:
\begin{equation}
\label{Eq:Margin}
p(\bm{\eta} | \mathbf{\delta t}) \propto \frac{\exp\left(-\frac{1}{2}\mathbf{\delta t}^T \bf{C}^{-1} \mathbf{\delta t}\right)}{\sqrt{\mathrm{det}(2\pi \bf{C})}} \times p(\bm{\eta}),
\end{equation}
where $\bf{C} = \bf{N} + \bf{T} \bf{B} \bf{T}^T$. In practice, the Woodbury matrix identity \citep{w50} is used to reduce Eq.\ (\ref{Eq:Margin}) to lower rank operations and thus accelerate computations.

\section{Destroying signal covariance}
\label{sec:kill_corr}
To assess GWB signal significance, we ideally want many equally-likely realizations of noise-only pulsar-timing datasets. If our detection statistic from the real dataset were smaller than in a fraction $\tilde{p}$ of these noise-only datasets, then the $p$-value of the GWB detection is less than or equal to $\tilde{p}$. Our aim in the following is to make this statement by removing correlations from the real pulsar timing datasets, instead of making many noise simulations where the properties are based on a potentially incomplete noise model. We try to make our noise models as realistic as possible, but any undiagnosed features in the data will not be represented in a noise simulation. Our approach keeps all features of the data intact, and we show that this produces a more conservative estimate of the GWB detection significance. Some progress has recently been made towards forming these kinds of null streams for continuous-GW analysis \citep{z+15,hl16}.

The reduced rank description of the GW signal covariance (as in Sec.\ \ref{sec:pta_likelihood}) provides two ways for us to remove the GWB's correlated influence between pulsars. The time-domain covariance induced by a GWB takes the following form:
\begin{align}
\label{Eq:ReduceRank}
\mathbf{C}_\mathrm{gwb} &= \langle \mathbf{F}\, \mathbf{a}_\mathrm{gwb} \mathbf{a}^*_\mathrm{gwb}\, \mathbf{F}^T \rangle  \nonumber \\
&= \mathbf{F}\, \langle \mathbf{a}_\mathrm{gwb} \mathbf{a}^*_\mathrm{gwb}\rangle\, \mathbf{F}^T \nonumber \\
&= \mathbf{F} \bm{\varphi}_\mathrm{gwb} \mathbf{F}^T,
\end{align}
where $\mathbf{F}$ is the Fourier design matrix of the signal, and $\bm{\varphi}_\mathrm{gwb}$ is the variance of the zero-mean signal coefficients $\mathbf{a}_\mathrm{gwb}$. This variance is proportional to the power spectral density of the GWB-induced time delays:
\begin{align}
\label{Eq:PhiMat}
\bm{\varphi}_\mathrm{gwb} &= \langle \mathbf{a}_\mathrm{gwb} \mathbf{a}^*_\mathrm{gwb} \rangle \nonumber \\
&= \Gamma_{ab} \times \frac{A^2_\mathrm{gwb}}{12\pi^2} \frac{1}{T} \left(\frac{\nu_i}{1\mathrm{yr}^{-1}}\right)^{-\gamma} \mathrm{yr}^2.
\end{align}

\begin{figure}
\centering
\includegraphics[width=0.3\textwidth]{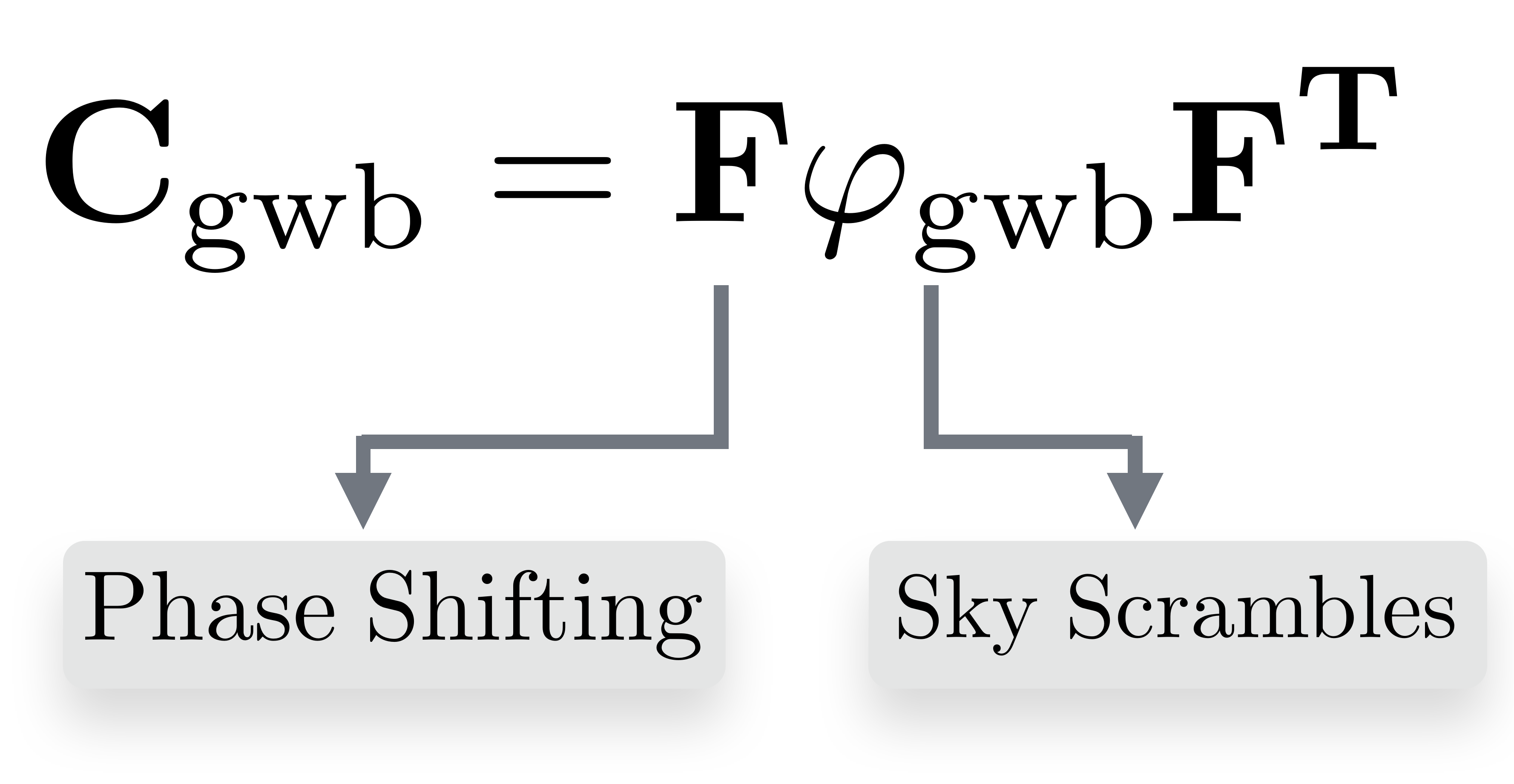} 
\caption{The two detection significance techniques operate on real pulsar timing datasets, and are naturally linked through the common result of destroying cross-pulsar GWB signal correlations.}
\label{figure:Unification}
\end{figure}

With the following two techniques we destroy signal covariance by either operating on the phase coherence through $\mathbf{F}$, or on the induced spatial correlations through $\Gamma_{ab}$. In Eq.\ (\ref{Eq:ReduceRank}) and Fig.\ \ref{figure:Unification} we see that these are naturally linked through the common result of mitigating cross-pulsar signal correlations in the data. 

\subsection{Phase shifting}
\label{Section:PhaseShifting}
Phase shifting attacks the phase coherence of the GWB signal that is induced between different pulsars in the PTA. There are two approaches one can take in phase shifting --- we can either construct phase shifted datasets (\textit{data-driven}), or we can search for the GWB with a phase shifted model (\textit{model-driven}). 

\textit{Data-driven}\,---\, In this approach we must first reconstruct the signal in each individual pulsar.  We determine the maximum likelihood parameters of the intrinsic pulsar noise (without including a GWB in the model), then solve Eq.\ (\ref{Eq:bmax}) to obtain the maximum likelihood signal coefficients, $\mathbf{\widehat{b}}$. By selecting only the components, $\mathbf{\widehat{a}}$, that correspond to the frequencies we want to shift, we can reconstruct the maximum likelihood signal realization with $\mathbf{s} = \mathbf{F}\mathbf{\widehat{a}}$.

We also construct a shifted signal, $\mathbf{s'}$, using the adjusted matrix $\mathbf{F'}$, defined as:
\begin{equation}
\label{Eq:FPrimeMatrix}
F'(\nu,t) = \sin\left(2\pi\nu t + \delta_\nu \right),
\end{equation}
and equivalent cosine terms, with $\delta_\nu$ a frequency-dependent random phase between 0 and $2\pi$. This gives our shifted signal as $\mathbf{s'} = \mathbf{F'}\mathbf{\widehat{a}}$.

We can then construct a new, shifted dataset $\mathbf{\delta t'}$:
\begin{equation}
\label{Eq:ShiftDataset}
\mathbf{\delta t'} = \mathbf{\delta t} - \mathbf{s} + \mathbf{s'}.
\end{equation}

\begin{figure}
\centering
\includegraphics[width=0.5\textwidth]{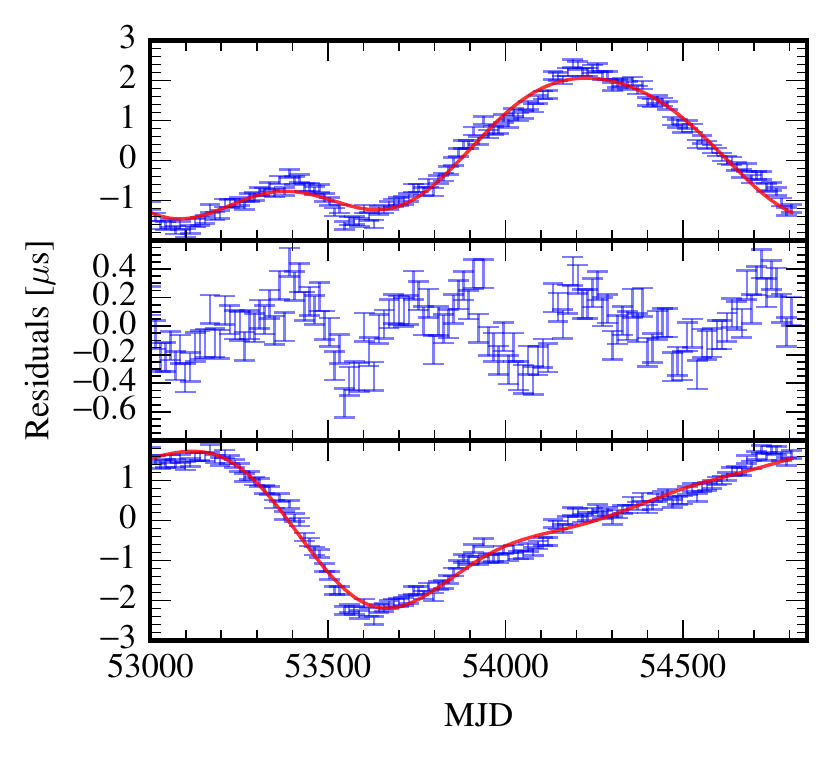}
\caption{Example of phase shifting on J$2317$$+$$1439$ in \textbf{Simulation 1} (blue points, top panel). The maximum-likelihood signal realization for the lowest $3$ frequencies is shown as a red line in the top panel.  We subtract this from the dataset to obtain a set of residuals (middle panel), where there is still clear structure left over. After phase shifting as in Section \ref{Section:PhaseShifting} we obtain the new signal (red line, bottom panel), which we add back to the residuals to obtain a new, shifted dataset (blue points, bottom panel).}
\label{figure:ExampleShuff}
\end{figure}

An example of this shifting process is shown in Fig.\ \ref{figure:ExampleShuff} for J$2317$$+$$1439$ in Simulation 1.\footnote{Residuals in the top panel of Fig.\ \ref{figure:ExampleShuff} are given by \textsc{Tempo2} performing a generalized least-squares fit of the timing-model parameters, where the \texttt{.par} file has the following red-noise estimates: $\mathrm{TNRedAmp}=-13.301$, $\mathrm{TNRedGam}=4.333$, $\mathrm{TNRedC}=50$.} Crucially (as will be shown in Sec.\ \ref{Section:Sim2}) this process retains the statistical properties of the original dataset, including any unmodelled stochastic or systematic effects. However by shifting the phases of the signal we have removed any correlations between pulsars.\footnote{We have implicitly assumed a stationary GWB signal, since our phase-shifting approach applies an independent random phase per sampling frequency. If the GW signal has non-stationary features then we need to account for frequency correlations in the data by performing correlated phase shifts, such that $\delta_i/\delta_j = \nu_i/\nu_j$. From here on we only consider stationary GW signals.}

\textit{Model-driven}\,---\, This approach is rather simple, in that we employ the pulsar datasets as they are, and search on the data with new phase-shifted Fourier design matrices, $\mathbf{F}'$, in the model for low-frequency processes. The data will now prefer a model with a common (but uncorrelated) red process instead of a GWB. We use the \textit{model-driven} phase-shifting in all of the following since it is so easily implemented. However, both approaches produce consistent results, and the \textit{data-driven} approach is important in illustrating that the statistical properties of each pulsar dataset remain unaffected.

\subsection{Sky scrambles}
\label{Sec:SkyScrambles}
Sky scrambles attack the \textit{spatial} correlations induced by the GWB signal. These correlations are described by the distinctive quadrupolar \textit{Hellings and Downs} signature. By contrast, a common uncorrelated low-frequency signal will have zero spatial correlations, and stochastic clock-standard drifts can be modeled as a low-frequency process with constant (monopole) spatial correlations \citep{hcm+12}. Inaccuracies in the Solar system ephemeris may lead to dipole-like spatial correlations between pulsars \citep{thk+15}, but modeling them as such is sub-optimal when coherent methods are available \citep{chm+10,dhy+13,ltm+15}. Unlike phase shifting, sky scrambling requires us to make specific assumptions about the expected spatial correlation signature of the signal. 

\textit{Model-driven}\,---\,
Our GWB search pipelines use a \textit{Hellings and Downs} template for the spatial correlations to filter out noise processes against the true signal. To sky scramble, we artificially move pulsar positions from their true values\footnote{Their true values are retained for fitting astrometric terms in the timing model.} so that the angular separations between pulsars will be scrambled. Thus, when we impose our template correlation signature it will be at odds with the spatial correlation signature of any true GWB in the data. Our goal is to make the overlap of the true spatial correlations as orthogonal as possible to our scrambled correlation model. We want to make it a maximally-poor template to effectively null the influence of GWB-induced spatial correlations in the data. We stress that we are not merely interchanging pulsar positions in the array --- the sky scrambles are constructed by searching for new pulsar positions over the entire sky to minimize the overlap of the scrambled template with the true correlations.

An intuitive picture is given by considering the behavior of the signal-to-noise ratio from the GWB optimal-statistic in the frequency domain:
\begin{equation}
\langle\rho\rangle^2 \propto \sum_{a, b\neq a} \sum_i \left[ \frac{ \Gamma'_{ab}(\nu_i)S'_i \times \Gamma_{ab}(\nu_i) S_i}{P_a(\nu_i) P_b(\nu_i)} \right],
\label{Eq:opt_stat}
\end{equation}
where $a$ and $b$ index pulsars, $i$ indexes sampling frequencies of the pulsar time series, primes denote template (or model) quantities, and unprimed quantities indicate true signal or noise processes in the data\footnote{In principle we can apply different template ORFs at each frequency, however for the purposes of this study we assume no frequency evolution of the GWB angular-power distribution, and thus no evolution of the true or template ORF.}. The power spectral density of the GWB-induced time delays, $S_i$, takes the usual power-law functional form throughout. This equation can be seen as a noise-weighted inner product of an ORF template with the actual correlations in the data.

Applying a sky-scrambled ORF to Eq.\ (\ref{Eq:opt_stat}) minimizes the overlap of the template with the signal, and diminishes the detection significance of the GWB in the data. This equation can also be used as a generator of sky scrambles --- we can insert typical pulsar noise properties along with signal assumptions to find a scrambled template ORF that makes $\langle\rho\rangle$ as small as possible. These scrambles can then be used to construct the spatial correlation template in a Bayesian analysis of real datasets. We repeat this process to generate many sky scrambles, and analyze the real data with the corresponding scrambled ORFs. Each analysis will return a Bayes factor for a GWB versus a common-uncorrelated red process. By virtue of the scrambling, these should now favor a model with a common-uncorrelated red process. The distribution of these Bayes factors is the desired null hypothesis distribution. We can then assess how frequently spurious noise correlations can give a Bayes factor that exceeds the Bayes factor found from the true unmodified dataset.

In practice, a more straightforward generator of sky scrambles is through minimization of the normalized inner product of the template ORF with the expected true ORF. This ``match statistic" has been explored in \citep{cs15b}, and we reiterate its form here:
\begin{align} \label{eq:match_stats}
M &= \frac{\sum_{a,b} \Gamma_{ab}\Gamma'_{ab}}{(\sum_{a,b} \Gamma_{ab}\Gamma_{ab} \times \sum_{a,b} \Gamma'_{ab}\Gamma'_{ab})^{1/2}}, \nonumber\\
\overline{M} &= \frac{\sum_{a, b\neq a} \Gamma_{ab}\Gamma'_{ab}}{(\sum_{a, b\neq a} \Gamma_{ab}\Gamma_{ab} \times \sum_{a,  b\neq a} \Gamma'_{ab}\Gamma'_{ab})^{1/2}},
\end{align}
where in $M$ the sum is over all unique pulsar pairings, while in $\overline{M}$ the sum excludes pulsar self-pairings since these merely add positive terms regardless of whether the pulsar positions are scrambled or not. The benefit of these match statistics is that they rely purely on the geometric properties of the array through the sky locations of the pulsars. We use the minimization of $\overline{M}$ to generate sky scrambles for the analyses in the rest of this paper, employing a particle swarm optimization (PSO) algorithm \citep{ke95,se98} to find scrambled positions for which $\overline{M}$ is below a given threshold with respect to the true ORF \textit{and} all other previously discovered sky-scrambles. 

There is a concern that, for a given number of pulsars, there are only a finite number of unique sky scrambles which produce ORFs that are orthogonal to the true ORF and all other scrambled ORFs. This is not easy to assess since we are not merely interchanging the pulsar positions, but it could bias our assessment of detection significance if there are repeated scrambles. Ideally we want all scrambles to be independent so that we have equally-weighted Bayes factors to produce the null hypothesis distribution. This is similar to the geometric problem known as a ``spherical code", where one tries to fit as many independent points as possible on the surface of a unit hypersphere whose position vectors have certain overlaps with each other. This issue is unsolved in an arbitrary number of dimensions, but if we were to insist on mutual-orthogonality of all scrambled ORFs then we can not have more than $N_\mathrm{psr}(N_\mathrm{psr}-1)$ scrambles, and even fewer if the scrambled ORFs must correspond to physical perturbations of pulsar positions. In practice we can partially mitigate this issue by not demanding that the scrambled ORFs be exactly mutually orthogonal, but merely that their normalized inner product be below some threshold. A larger threshold value gives more sky-scrambles, but at the cost of reduced independence.

\textit{Data-driven}\,---\, We perform an initial search on the entire pulsar array dataset for a GWB signal, from which we extract the maximum likelihood signal coefficients $\mathbf{\widehat{a}}_\mathrm{gwb}$ in each pulsar. At each frequency, the variance of these signal coefficients is equal to the power spectral density of the GWB-induced time delays, scaled by the ORF between the pair of pulsars in question: $\langle a_a a_b^*\rangle_i = \Gamma_{ab} S_i$. In the following we denote the vector of all pulsars' GWB signal coefficients at a particular frequency, $\nu_i$, by $\mathbf{\widehat{a}}_i$. The expected covariance matrix of this vector of coefficients is then the \textit{Hellings and Downs} spatial correlation matrix, scaled by $S_i$. Explicitly, the spatial correlation matrix, $\mathbf{\Gamma}$ is:
\begin{equation}
\mathbf{\Gamma} = \begin{bmatrix} \Gamma_{11} & \Gamma_{12} & \vdots \\ \Gamma_{21} & \Gamma_{22} & \vdots \\ \cdots & \cdots & \ddots \end{bmatrix}.
\end{equation}

We Cholesky factorize $\mathbf{\Gamma}$, then operate on $\bm{\widehat{a}}_i$ with the inverse Cholesky factor to decorrelate the signal between different pulsars. Hence,
\begin{align} \label{eq:scrambled_decorrelate}
\mathbf{\Gamma} &= \mathbf{L}\mathbf{L}^T, \nonumber\\
\mathbf{\widehat{a}''}_i &= \mathbf{L}^{-1}\mathbf{\widehat{a}}_i,
\end{align}
where $\mathbf{\widehat{a}''}_i$ is the vector of new pulsar signal coefficients at frequency $i$, which are uncorrelated between pulsars but retain the same spectral properties. We repeat this process at all sampling frequencies in our rank-reduced approximation of the GWB signal, giving new vectors of signal coefficients for each pulsar. However, this produces only one set of uncorrelated signal coefficients --- to produce many scrambled datasets we can correlate the coefficients again by Cholesky factorizing a scrambled ORF, $\mathbf{\Gamma'}=\mathbf{L'}\mathbf{L'}^T$, and operating on the uncorrelated signal coefficients such that $\mathbf{\widehat{a}'}_i = \mathbf{L'}\mathbf{\widehat{a}''}_i$. As in the \textit{data-driven} phase shifting approach, we now form new pulsar datasets such that:
\begin{equation}
\mathbf{\delta t'} = \mathbf{\delta t} - \mathbf{F}\mathbf{\widehat{a}} + \mathbf{F}\mathbf{\widehat{a}'}.
\end{equation}

We now have new pulsar datasets with their individual spectral properties intact, but which are correlated according to a scrambled ORF. Each scrambled ORF gives a new PTA dataset which is analyzed under the assumption that the true ORF is present, and iterating over scrambles gives the distribution of the Bayes factor under the null hypothesis. The analog with Eq\@.~(\ref{Eq:ShiftDataset}) is now easy to see: when phase shifting we modify $\mathbf{F}$ while in sky scrambling we modify $\mathbf{\widehat{a}}$. As in the case of phase shifting, the \textit{model-driven} sky-scrambling approach is a more straightforward practical implementation, so we use it in all of the following.

\subsection{Unified formalism}
We now examine the combined influence of phase shifting and sky scrambling on the timing-residual correlation between two pulsars. For simplicity we consider only the correlation due to the GWB. We also initially consider only one sampling frequency in the reduced rank description of the signal, but generalize later. The covariance between timing residuals at $t_{ak}$ in pulsar $a$ and at $t_{bl}$ in pulsar $b$ is:
\begin{equation}
\label{Eq:CovRedRank}
\mathbf{C}_{(ak),(bl)} = \mathbf{F}_{(ak)} \bm{\varphi}_{ab} \mathbf{F}_{(bl)}^T,
\end{equation}
where, with only one sampled frequency at $\nu_i$, $\mathbf{F}_{(ak)}$ takes the form:
\begin{equation}
\mathbf{F}_{(ak)} = \begin{bmatrix} \sin(2\pi\nu_i t_{ak}) & \cos(2\pi\nu_i t_{ak}) \end{bmatrix},
\end{equation}
and $\mathbf{F}_{(bl)}$ is likewise. The spectrum, $\varphi_{ab}$, at each frequency is as in Eq.\ (\ref{Eq:PhiMat}), but is explicitly represented here as a $2N_\mathrm{freqs} \times 2N_\mathrm{freqs}$ matrix (since each frequency has a sine and cosine basis function):
\begin{equation}
\bm{\varphi}_{ab} = \begin{bmatrix} \varphi_{ab} & 0 \\ 0 & \varphi_{ab} \end{bmatrix}.
\end{equation}

The result of phase shifting and sky scrambling is to convert $\mathbf{F}_{(ak)}$ and $\varphi_{ab}$ into the following:
\begin{align}
\mathbf{F}'_{(ak)} &= \begin{bmatrix} \sin(2\pi\nu_i t_{ak} + \delta_{ai}) & \cos(2\pi\nu_i t_{ak} + \delta_{ai}) \end{bmatrix}, \nonumber\\
\varphi'_{ab} &= \frac{\Gamma'_{ab}}{\Gamma_{ab}} \times \varphi_{ab}.
\end{align}

If we explicitly evaluate Eq.\ (\ref{Eq:CovRedRank}) with the phase shifted and scrambled quantities (and generalize to multiple sampling frequencies) we get the following for our scrambled model of the induced correlations:
\begin{align} \label{eq:scramphase_cov}
\mathbf{C}'_{(ak),(bl)} = \displaystyle\sum^{N_\mathrm{freqs}}_i \frac{\Gamma'_{ab}}{\Gamma_{ab}} &\varphi_{ab} \left[ \cos(2\pi\nu_i(t_{ak}-t_{bl}))\cos(\delta_{ai}-\delta_{bi}) \right. \nonumber\\ 
&\left. - \sin(2\pi\nu_i(t_{ak}-t_{bl}))\sin(\delta_{ai}-\delta_{bi}) \right].
\end{align}

One can easily see that without phase shifting (or with a common phase shift for all pulsars at each frequency) and without sky scrambling, the correlation is:
\begin{equation} \label{eq:wk_theorem}
\mathbf{C}'_{(ak),(bl)} = \displaystyle\sum^{N_\mathrm{freqs}}_i \varphi_{ab} \cos(2\pi\nu_i(t_{ak}-t_{bl})),
\end{equation}
which is the just the discrete Wiener-Khinchin conversion between the power spectral density of a process and the time-domain correlation. The autocovariance of each pulsar can be examined by setting $a=b$ in Eq.\ (\ref{eq:scramphase_cov}). In this case $\Gamma_{aa}=\Gamma'_{aa}=1$, and phase shifts cancel at each sampling frequency, such that the statistical properties of each individual pulsar dataset remain intact.

\section{Application to simulations}
\label{sec:sims}
In the following we test our two techniques against several different types of PTA datasets. All Bayesian analysis and evidence recovery is performed using the PTA analysis suite \textsc{NX01} \citep{nx01} with the \textsc{MultiNest} sampler \citep{fh08,fhb09,fhc+13,bgn+14}, where $1000$ live points are employed in Secs.\ \ref{Section:Sim1} and \ref{Section:Sim3} (for a $2$-D parameter space), and $5000$ live points are employed in Secs.\ \ref{Section:Sim2} and \ref{Section:Sim4} (for a $22$-D and $26$-D parameter space, respectively).

\begin{figure*}
\centering
\subfloat[\textbf{Simulation $1$} \label{fig:shuffsimfreqs1}]{\includegraphics[width=0.5\textwidth]{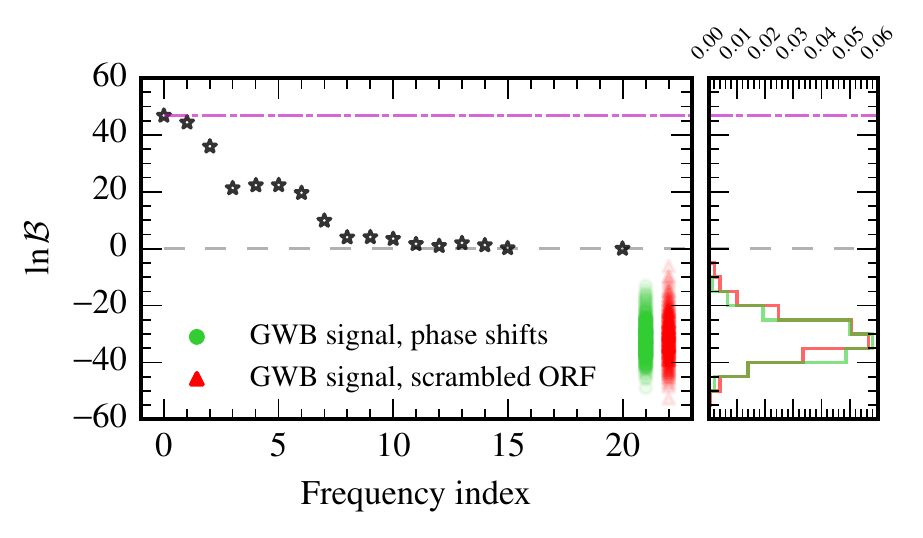}}
\subfloat[\textbf{Simulation $2$} \label{fig:shuffsimfreqs2}]{\includegraphics[width=0.5\textwidth]{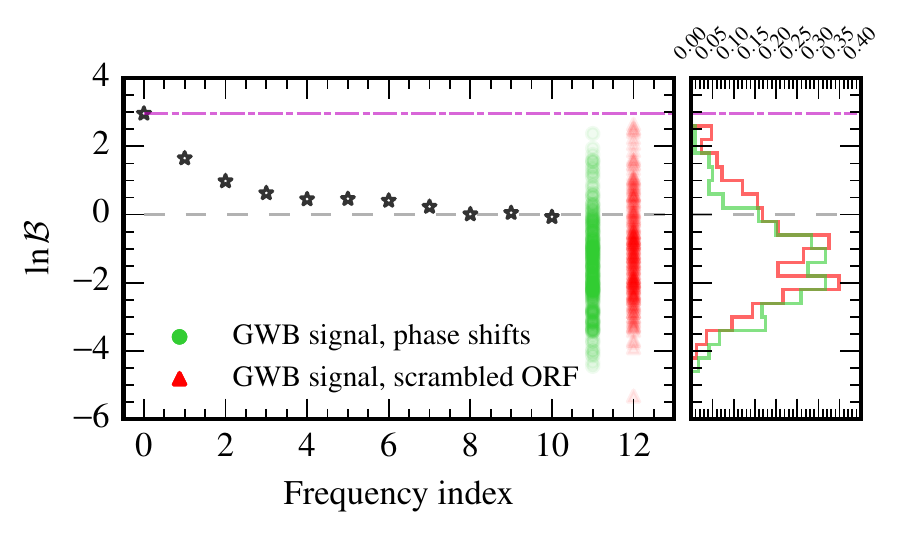}}
\caption{Log-Bayes factor for a correlated GWB model versus a common-uncorrelated red process. Black stars correspond to operating on the real dataset without modeling spatial correlations in the lowest $N$ coefficients, e.g.\ the black star at a frequency index of $1$ signifies that the correlation has not been included in the lowest frequency. The green circles and histogram correspond to operating on the unmodified dataset with $300$ phase-shift instances, while the red triangles and histogram are for $300$ sky-scramble instances. The case of equal model evidences is indicated with a dashed grey line at zero, and the true Bayes factor of spatial correlations in the dataset is indicated with a dash-dot magenta line.}
\label{figure:Shift1Sim}
\end{figure*}

\begin{figure}
\centering
\subfloat[\textbf{Simulation $1$} \label{fig:scramphasesim_sim1}]{\includegraphics[width=0.5\textwidth]{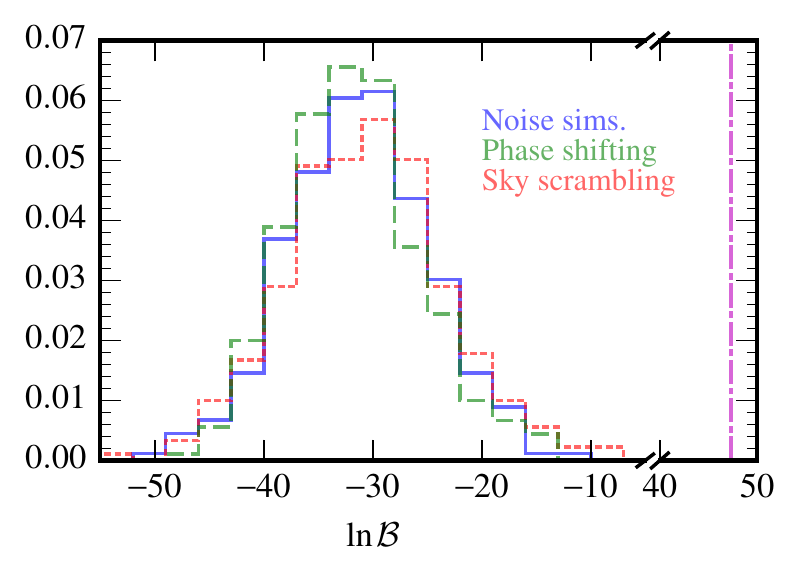}}\\
\subfloat[\textbf{Simulation $2$} \label{fig:scramphasesim_sim2}]{\includegraphics[width=0.5\textwidth]{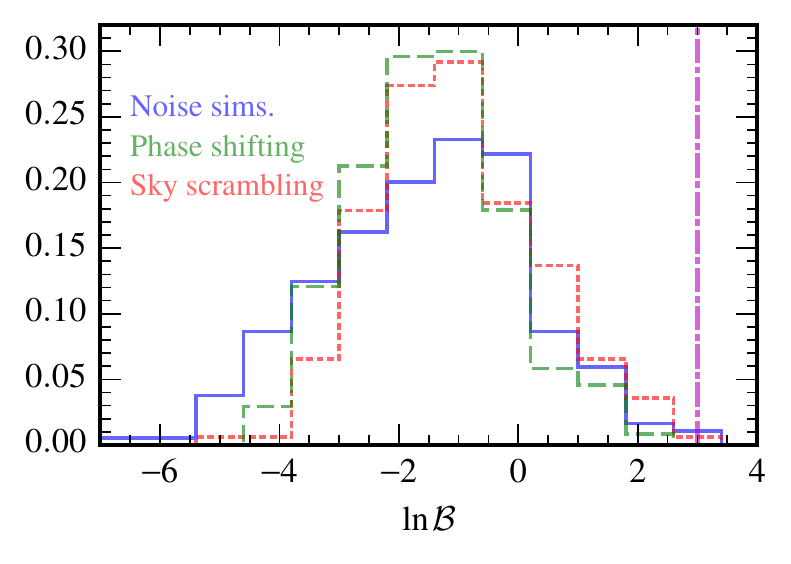}}
\caption{The null hypothesis distribution of Bayes factors for a GWB versus a common-uncorrelated red process is obtained with different techniques. The result of $300$ phase shifts (green, long-dash) and $300$ sky scrambles (red, short-dash) on the real dataset are compared to the distribution from noise simulations (blue, solid). The Bayes factor from the real datasets are shown as a vertical dash-dot magenta line in both panels.}
\label{figure:ScrambleBFdist}
\end{figure}

\subsection{Simulation 1 (IPTA MDC open 1)}
\label{Section:Sim1}
We apply the techniques to the first open dataset of the publicly available, International Pulsar Timing Array \citep{v+16} mock data challenge.\footnote{\url{http://www.ipta4gw.org/?page_id=214}}  This dataset contains 5 years of observations for a set of 36 pulsars, each with a 14 day cadence, and each with uncorrelated TOA measurement uncertainties of $10^{-7}$ seconds.  These datasets do not include any intrinsic red noise processes, clock errors, or Solar system ephemeris uncertainties, but do include a GWB with a power-law spectrum ($A_{\mathrm{gwb}} = 5\times10^{-14}$, $\gamma_\mathrm{gwb}= 13/3$).  While this does not represent a realistic dataset by any metric, it is a simple initial test case upon which to explore the effectiveness of our approaches for eliminating the correlation between pulsars due to the GWB. In the following analyses, our noise model includes only TOA uncertainties given by the observations (i.e.\ no search parameters) while the signal model is a power-law GWB (two search parameters).

We first need to assess how many sampling frequencies contain information about correlations between the pulsars, and thus how many frequencies we need to phase shift. We analyze the dataset with a model which neglects spatial correlations in a successive number of sampling frequencies, beginning with the base frequency $1/T$ and then increasing. As we see in Fig.\ \ref{fig:shuffsimfreqs1}, the evidence for a GWB without spatial correlations in the lowest $\sim 20$ frequencies is indistinguishable from a common-uncorrelated red process. So long as we apply random phase shifts to at least the first $20$ sampling frequencies in our model, the phase coherence between pulsars will be destroyed. Sky scrambles are generated using the match-statistic minimization approach described in Sec.\ \ref{Sec:SkyScrambles}, with a threshold of $0.2$. 

The result of carrying out $300$ phase-shifting and $300$ sky-scrambling analyses of the dataset are also shown in Fig.\ \ref{fig:shuffsimfreqs1} as green and red log-Bayes factor histograms, respectively. The Bayes factor is for a GWB versus a common-uncorrelated red process. The fact that these histograms are centred around $\sim$$-30$ shows two things: $(i)$ the data contains a lot of information about spatial correlations; and $(ii)$ the data now strongly favor a common-uncorrelated red process under the modified models. The phase-shifting and sky-scrambling techniques produce consistent null hypothesis (i.e. no correlations) distributions. The true log-Bayes factor of $\sim +47$ is seen to be highly significant in the context of these distributions. To make robust statements about the $p$-value of correlations one would need to perform many more analyses than our $300$ to fill out the tails of the distributions. However it is clear that spurious noise correlations are an improbable source of producing the very high Bayes factor given by the data, with a probability of $< 1/300$. Indeed, this is a clear-cut case to begin with since the Bayes factor already favors the signal model by $e^{47}:1$ odds. We will see the real use of these techniques in the more realistic and marginal simulation to follow.

Finally, we check how phase-shifting and sky-scrambling compare to noise simulations as a way to construct the null hypothesis Bayes factor distribution. In Fig.\ \ref{fig:scramphasesim_sim1} we compare our histograms of log-Bayes factors from phase-shifting and sky-scrambling with a histogram from analyzing $300$ independent noise-only datasets. In the latter, each pulsar has the same statistical properties as in the original dataset but without spatial correlations. We see that our techniques match the performance of noise simulations very well.

\subsection{Simulation 2 (a more realistic simulation)}
\label{Section:Sim2}
We now apply the techniques to a more realistic $10$-pulsar EPTA dataset.  Table \ref{Table:Sim2details} lists the timespan, rms of the white noise, and the properties of the red noise for each pulsar in the simulation. These values are chosen to be similar to those given in \citet{cll+15} in order to provide as realistic a simulation as possible. The observation schedule matches that of the true EPTA pulsars. We also add a power-law GWB signal with spectral index, $\gamma_{\mathrm{gwb}} = 13/3$, and amplitude $A_{\mathrm{gwb}} = 5\times10^{-15}$.  While this is significantly in excess of current upper limits, we choose this amplitude so that the change in log-evidence between models that do or do not include spatial correlations is $\sim$$+3$. Therefore the phase-shifting and sky-scrambling operations can produce a measurable change in the evidence.
\begin{table}
\centering
\caption{Details of simulation 2.} 
\centering 
\begin{tabularx}{0.5\textwidth}{b b b b s} 
\hline\hline 
Pulsar  & $T_\mathrm{obs}$~[years] & $\sigma_w$~[$\mu$s]	& $\log_{10}A_{\mathrm{red}}$ & $\gamma_{\mathrm{red}}$ \\
\hline 
J$0613$$-$$0200$ & 	16.054	&	1.58    &	-13.90		&	3.18		\\
J$0751$$+$$1807$ & 	17.606	&	2.60    &	-14.14		&	2.58		\\
J$1012$$+$$5307$ &  	16.831	&	1.47    &	-13.09		&	1.65		\\
J$1640$$+$$2224$ & 	16.735	&	1.99    &	-13.24		&	0.03		\\
J$1643$$-$$1224$ &  	17.300	&	1.65    &	-18.56		&	4.04		\\
J$1713$$+$$0747$ & 	17.657	&	0.26    &	-14.90		&	4.85		\\
J$1744$$-$$1134$ &  	17.250	&	0.65    &	-13.60		&	2.00		\\
J$1857$$+$$0943$ & 	17.310	&	1.51    &	-16.00		&	1.35		\\
J$1909$$-$$3744$ & 	9.379	&	0.12    &	-13.99		&	2.06		\\
J$2145$$-$$0750$ & 	17.161	&	1.19    &	-13.87		&	4.02		\\
\hline
\end{tabularx}
\label{Table:Sim2details} 
\end{table}

As in the previous section, we first investigate how many frequencies are informative of spatial correlations in the data. We see in Fig.\ \ref{fig:shuffsimfreqs2} that the majority of information is contained in the lowest $\sim 2-3$ frequencies. The evidence is reduced to that of a common-uncorrelated red process after neglecting spatial correlations in the lowest $\sim 10$ frequencies. We therefore apply phase shifts to at least these lowest $10$ frequencies in the GWB signal model. As before, sky scrambles are generated by minimizing the unique off-diagonal match statistic, $\overline{M}$, with a threshold of $0.5$.

Figure \ref{fig:shuffsimfreqs2} also shows the histograms of log-Bayes factors produced from several hundred phase-shifting and sky-scrambling experiments, where the techniques are shown to match very well. Although not as significant as the signal in IPTA MDC open 1, we see that the Bayes factor for spatial correlations in this more realistic simulation is still highly convincing and unlikely to have been formed via spurious noise correlations. As discussed previously, in a real analysis we would desire a quantitative assessment of the significance; this would require many more phase shifting or sky scrambling experiments than are examined here in order to produce smooth distributions which are well sampled in the tails. We must also bear in mind that we used a larger match threshold than before to generate the sky-scrambles. Therefore they are not completely independent of one another, which may introduce some bias in assessments of detection significance (see Sec.\ \ref{Sec:SkyScrambles}).  

As before, we confirm that phase-shifting and sky-scrambling produce Bayes factor distributions under the null hypothesis which compare well with the distribution produced from analyzing many noise-only simulations. The results for this are shown in Fig.\ \ref{fig:scramphasesim_sim2}, where all distributions are shown to be in good agreement.

\begin{figure}
\centering
\includegraphics[width=0.5\textwidth]{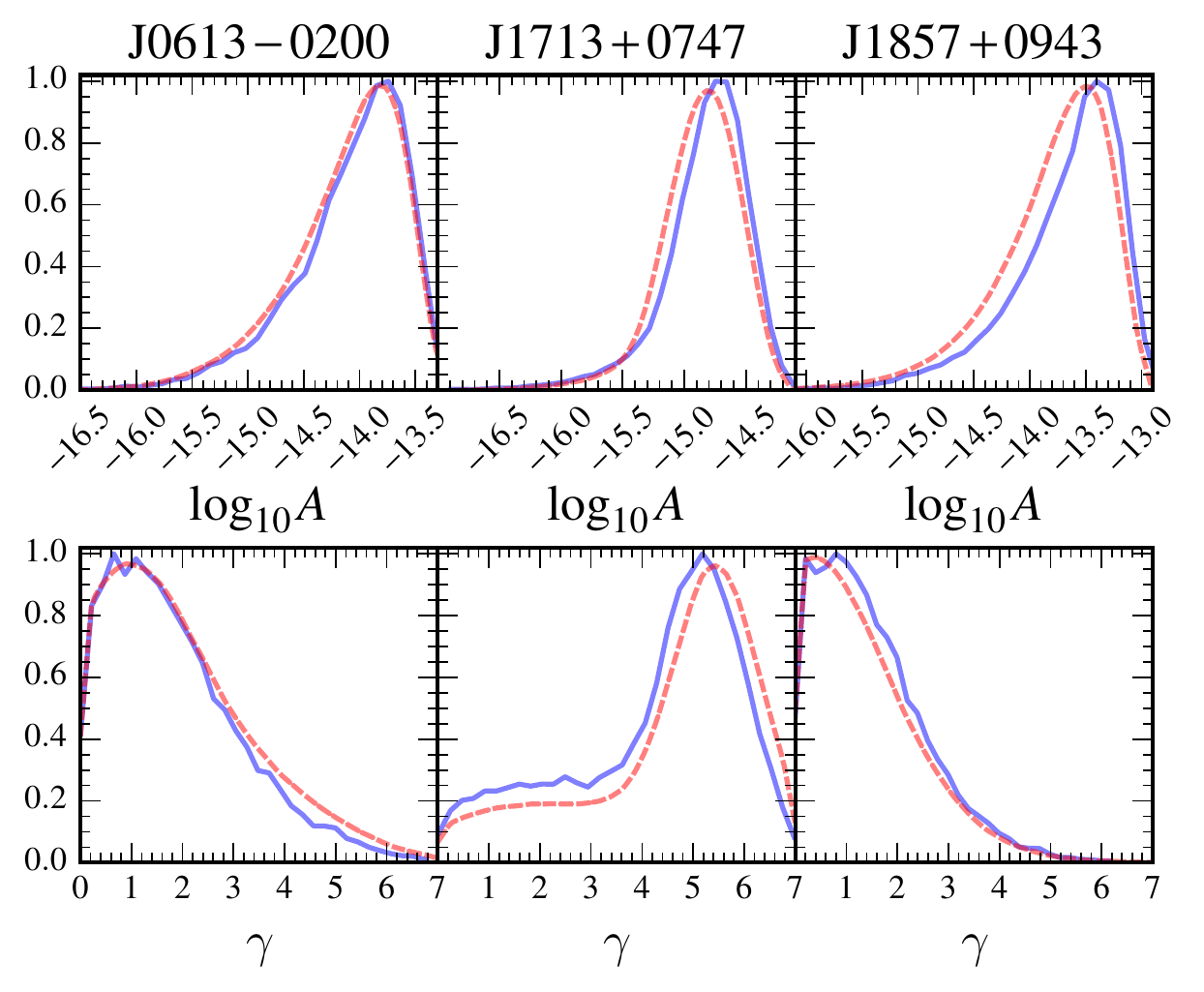} 
\caption{One dimensional marginalized posterior distributions for the log-amplitude (top) and spectral index (bottom) of the intrinsic red noise for three of the pulsars in \textbf{simulation 2}.  Blue lines represent the analysis for the unmodified dataset containing the uncorrelated common red noise term, red lines are the mean of the posterior distributions over $50$ realizations of the combined phase and position shifted datasets.}  
\label{figure:ParamComps}
\end{figure}

Finally, we use this more realistic dataset to demonstrate that phase-shifting and sky-scrambling do not alter the statistical properties of each individual pulsar dataset. Figure \ref{figure:ParamComps} shows the posterior distributions of the intrinsic red noise parameters for a subset of the pulsars in simulation 2. Red lines show the mean parameter estimates over $50$ realizations of simultaneous phase-shifting and sky-scrambling on the original dataset, where the model includes separate red noise per pulsar and a GWB. Blue lines show parameter estimates from an analysis of the unmodified dataset, where the model includes separate red noise per pulsar and a common-uncorrelated red process. As desired, the shifting and scrambling processes have not significantly affected the parameter estimates for individual pulsars. This indicates that the underlying statistics of the dataset remain consistent whether we analyze with a shifted/scrambled GWB model or with a common-uncorrelated red process.

\subsection{Influence of unmodeled noise processes}
So far we have shown that phase-shifting and sky-scrambling provide null hypothesis distributions that agree with what is given by noise simulations. We now go further to show that phase-shifting and sky-scrambling can in fact be superior. This occurs when our noise models are poor so that our understanding of noise processes in the dataset is incomplete. 

\subsubsection{Simulation 3 (mismodeled intrinsic noise)}
\label{Section:Sim3}

We generate an extreme example of a dataset with large unmodeled noise that is intrinsic to each pulsar. We inject loud independent glitches (negative ramps in the residual time-series due to spontaneous increases in the pulsar rotational frequency) into each pulsar of simulation 1 (described in Sec.\ \ref{Section:Sim1}). This glitch term is injected as:
\begin{equation} \label{eq:glitchsignal}
s_\mathrm{glitch}(t) = -\mathcal{A}\times(t - t_e)H(t - t_e) \times\mathrm{spd},
\end{equation}
where $t$ is the MJD of a given pulsar observation, $H(\cdot)$ is the Heaviside step function, $\mathcal{A}$ is the glitch amplitude, $t_e$ is the MJD of the glitch epoch, and $\mathrm{spd}=86400$ is the number of seconds per day. The glitch epochs are randomly drawn as $t_e\in \mathrm{MJD}\;U[53000,54806]$, while the glitch amplitudes are randomly drawn as $\log_{10}\mathcal{A}\in U[-18,-17]$. 

We analyze this simulation as we would do for a real dataset, with the following checklist:
\begin{enumerate} \itemsep0em 
\item Analyze the true PTA dataset for a GWB signal.
\item Analyze the true PTA dataset for a common-uncorrelated red process.
\item Generate many noise-only simulated datasets. We simulate pulsars with the maximum-likelihood noise properties of the true dataset, and also inject a common-uncorrelated red process with the maximum-likelihood parameters of the recovered GWB signal in the true dataset. Analyze each simulated dataset for a GWB signal and for a common-uncorrelated red process.
\item Analyze the true dataset to assess how many sampling frequencies contain information about spatial correlations. Perform many phase-shift analyses on the true dataset.
\item Generate scrambled sky positions from either an optimal-statistic analysis or the match statistics. Perform many sky-scramble analyses on the true dataset.
\item Item $(3)$ gives the distribution of the Bayes factor under the null hypothesis from simulations. Items $(4)+(2)$ are used to give the phase-shifting estimation of the null hypothesis Bayes factor distribution. Items $(5)+(2)$ are used to give the sky-scrambling estimation of the null hypothesis Bayes factor distribution. 
\end{enumerate}

For item $(1)$, our model consists of a GWB signal, and white-noise given by the reported TOA measurement uncertainties. We purposefully do not model the glitch in each pulsar to assess how phase-shifting and sky-scrambling perform when we have an incomplete noise model. The parameter estimation of the GWB spectrum will be dominated by the self-pairings of the pulsars (since the off-diagonal elements of the ORF matrix are at most half of the diagonal terms). Therefore the glitches (which are negative ramps) are interpreted as an additional low-frequency component of the GWB, so that the recovered signal will have a higher amplitude than the true signal. This is indeed the case, where the maximum-likelihood GWB signal parameters are found to be $A_\mathrm{gwb}=9.60\times 10^{-14}$ and $\gamma_\mathrm{gwb}=4.24$, and the posterior distributions are inconsistent with the true GWB signal parameters of $A_\mathrm{gwb}=5\times 10^{-14}$ and $\gamma_\mathrm{gwb}=4.33$. The log-Bayes factor for spatial correlations in this dataset is $\sim +4$.

\begin{figure}
\centering
\includegraphics[width=0.5\textwidth]{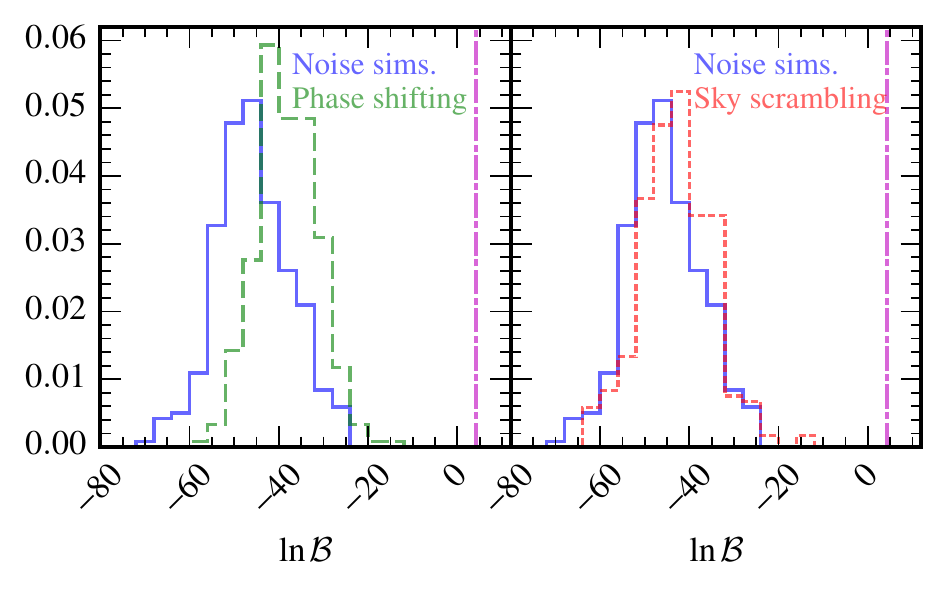}
\caption{\textbf{Simulation 3:} Null hypothesis distributions of Bayes factors for a GWB versus a common-uncorrelated red process. \textbf{Left:} performing $300$ noise simulations (blue solid, both panels) is compared to performing $300$ phase-shifting experiments (green long-dash, left panel). \textbf{Right:} the noise simulations are compared to performing $300$ sky-scrambling experiments (red short-dash, right panel). The Bayes factor in the true dataset is shown as a vertical dash-dot magenta line in both panels.}
\label{fig:sim3_all}
\end{figure}

We create $300$ PTA dataset simulations which contain white-noise at the level of the reported TOA uncertainties, and a common-uncorrelated red process in each pulsar with $A=9.60\times 10^{-14}$ and $\gamma=4.24$. Since this amplitude is larger than what is actually in the true dataset, the significance of the GWB signal should be biased high. The histogram of log-Bayes factors from these noise simulations is contrasted with the phase-shifting histogram and sky-scrambling histogram in Figure \ref{fig:sim3_all}. Phase-shifting gives a more conservative estimate of the GWB significance in the true dataset (dash-dot magenta line) than noise simulations, while sky-scrambling gives a significance that is less than noise simulations but more than phase-shifting.

\subsubsection{Simulation 4 (influence of clock and ephemeris errors)}
\label{Section:Sim4}

\begin{figure}
\centering
\includegraphics[width=0.5\textwidth]{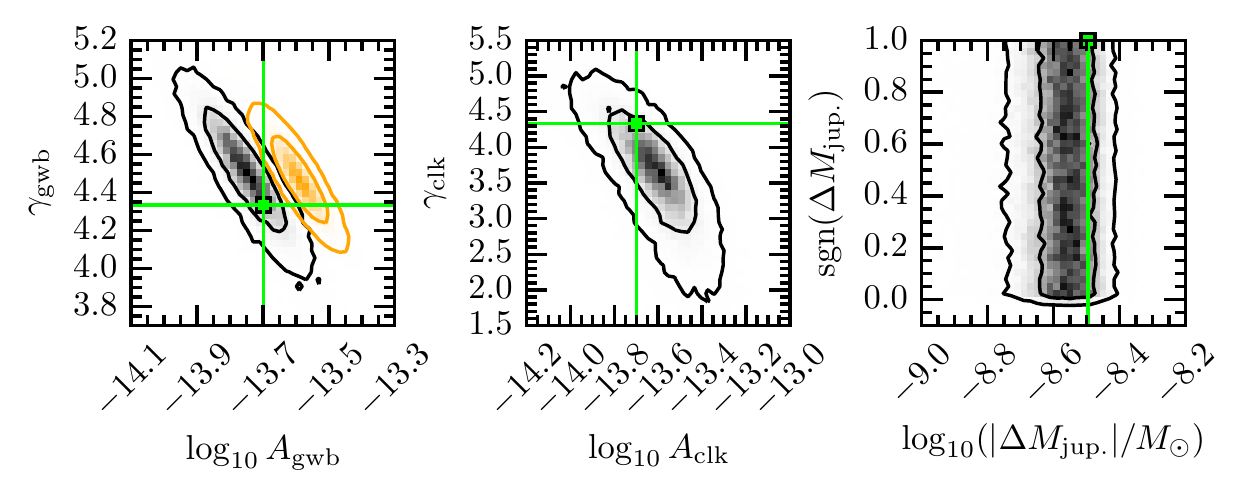}
\caption{\textbf{Simulation 4:} $2$-dimensional marginalized posterior distributions for GWB spectrum parameters (left panel), clock-error spectrum parameters (middle panel), and ephemeris uncertainty parameters (Jupiter mass error, sign of mass error; right panel). Black lines mark the $68\%$ and $95\%$ credible regions when all processes are modeled, while the orange corresponds to only a GWB being modeled. Light green lines and points indicate the injected parameters.}
\label{fig:sim_gwbephemclock}
\end{figure}

\begin{figure}
\centering
\includegraphics[width=0.5\textwidth]{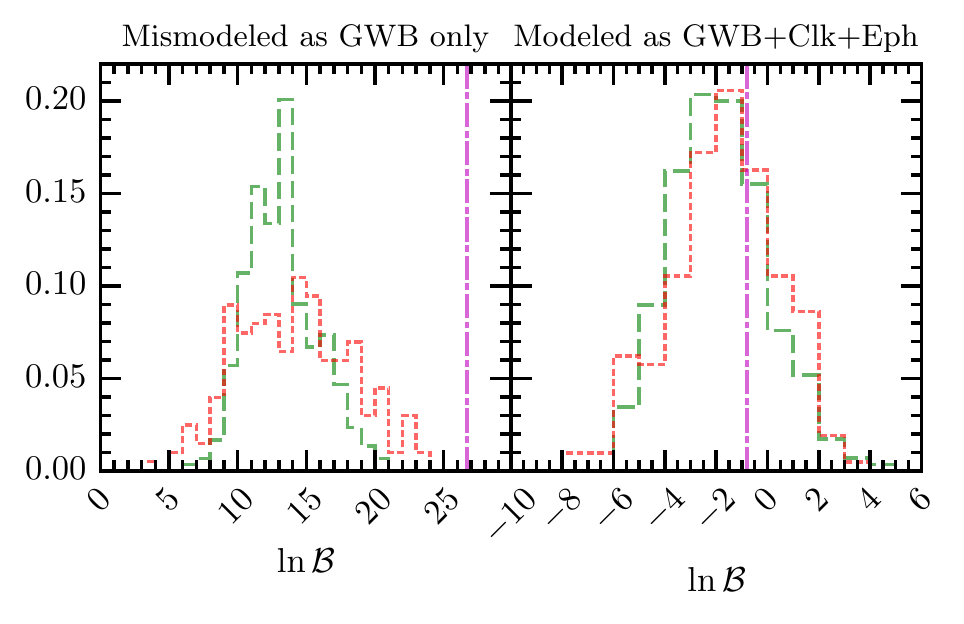}
\caption{\textbf{Simulation 4:} Null hypothesis distributions of Bayes factors for a GWB versus a common-uncorrelated red process. \textbf{Left:} Phase-shifting (green long-dash) and sky-scrambling (red short-dash) diagnose the presence of other spatially-correlated signals in this dataset, since the null hypothesis distributions still favor a GWB. \textbf{Right:} If we correctly include clock and ephemeris errors in our noise model, then phase-shifting and sky-scrambling give consistent null hypothesis distributions, showing that GWB correlations in this dataset are insignificant.}
\label{fig:sim2_gwbvsdiag_PhaseSim_gwbclkephem}
\end{figure}

As mentioned previously, a GWB is not the only process that can induce spatial correlations between pulsars. We create a new dataset with the same realistic properties as simulation 2, and inject a power-law GWB signal with spectral index, $\gamma_{\mathrm{gwb}} = 13/3$, and amplitude $A_{\mathrm{gwb}} = 2\times10^{-14}$. This signal is four times larger in amplitude than in simulation 2. We also add two other processes that could induce spatial correlations --- $(i)$ a stochastic clock-error process that has the same waveform across all pulsars, and has the same power-law spectrum parameters as the GWB; $(ii)$ a very large error in the mass of Jupiter in our model of the Solar system ephemeris ($\Delta M_\mathrm{jup.} = 3.2\times 10^{-9}M_\odot$). 

We model the clock error as a stochastic process with monopole correlations between pulsars (the correlation is $1$ at all angular separations). Previous work has noted that clock errors only induce monopole correlations if the pulsar data spans are identical and the same timing model fits are applied \citep{hcm+12}. However our Bayesian approach performs regression on the timing model simultaneously with all noise and signal processes. So we do not need to be concerned about potential inference biases caused by the timing model fit. Furthermore, we model the ephemeris uncertainty as a coherent process rather than as a spatially-correlated signal. We search for the magnitude and sign of the Jupiter mass error \citep{chm+10}. If an ephemeris uncertainty is not simply due to a planetary mass offset then we can still model the uncertainty coherently \citep{ltm+15,dhy+13}, without needing to resort to a dipole-like spatial-correlation model.\footnote{In this approach, we marginalize over the separate components of an ephemeris-error vector with zero-mean Gaussian priors, whose variances we parametrize. The variance is proportional to the power-spectrum of the ephemeris-error component, and can be modeled with a power-law or free-spectrum. If ephemeris uncertainties are caused by a large number of objects (like asteroids) then they should show up as spikes in the recovered spectrum at their relevant orbital frequencies. This approach is still coherent because we retain directional information of the ephemeris uncertainties.}

Figure \ref{fig:sim_gwbephemclock} shows the $2$-D marginalized posterior distributions for the GWB, clock error, and Jupiter mass errors from analyzing this simulation. The black credible regions correspond to the case where all processes are simultaneously modeled, while the orange credible regions are when the simulation is assumed to contain only a GWB. When all processes are modeled, the recovered posteriors are consistent with the injected values. We have verified that either the clock error or ephemeris uncertainty by themselves would still produce systematic parameter-estimation errors if unmodeled. 

In Fig.\ \ref{fig:sim2_gwbvsdiag_PhaseSim_gwbclkephem} we phase-shift and sky-scramble on this simulation with different noise model assumptions. In the left panel, we assume that the noise is characterized by the TOA uncertainties and intrinsic pulsar red-noise. In the right panel the noise model additionally includes a clock error and Jupiter mass error. In the left panel we now see the limitations of phase-shifting and sky-scrambling, since the clock-error produces a large positive offset in the spatial correlations between pulsars. Referring back to Eq.\ (\ref{eq:scramphase_cov}), we see that phase-shifting essentially multiplies the spatial correlations in the true dataset by a random number in the range $[-1,1]$. Although this makes \textit{Hellings \& Downs} correlations consistent with zero, this does not work when we have spatial correlations with a large positive offset. Model selection still prefers a GWB signal over a common-uncorrelated red process. Likewise, model selection still prefers our scrambled ORF over an uncorrelated model. 

These limitations make phase-shifting and sky-scrambling very valuable diagnostic tools --- the fact that they could not completely eliminate all spatial correlations is an indication of an unmodeled spatially-correlated process. When we assume that only a GWB is present, and make noise simulations to assess detection significance, we get a highly significant (biased) detection of a GWB. By contrast, in the right panel of Fig.\ \ref{fig:sim2_gwbvsdiag_PhaseSim_gwbclkephem} we include a clock error and Jupiter mass error in our noise model. The GWB is then properly isolated from these two effects. We only phase shift in the GWB basis functions, leaving the clock spatial correlations unaffected. Likewise, only the GWB correlation signature is sky-scrambled. When we do this we get consistent null hypothesis distributions which show that the GWB spatial correlations are insignificant in this dataset. The spatial correlations are swamped by the clock error. Hence our techniques are valuable diagnostic tools, and can be used in conjunction with models for other spatially-correlated processes to properly isolate the significance of the GWB.

\section{Conclusions}
\label{sec:conclusions}
We have studied two methods of determining the significance of a GWB in a pulsar-timing array. Within our Bayesian context, we can compute the Bayes factor for a spatially-correlated GWB signal versus a common-uncorrelated red process in the pulsars. But to put this in context we need to know how often spurious noise correlations can give similar Bayes factors. If noise alone can often produce values as large as what we see in the true data then this Bayes factor is clearly not very significant. Standard rule-of-thumb guides exist for assessing Bayes factor significance \citep{j61,kr95}, but using them in any production-level analysis is unsatisfactory since they are not designed with the specifics of a given problem in mind. We must resort to numerical experiments to produce distributions of Bayes factors under the null hypothesis i.e.\ where there are no spatial correlations in the dataset. 

Our first technique involves adding random phase shifts to all basis functions modeling the low-frequency processes in a given pulsar dataset. This is performed separately for each pulsar so that the statistical properties of each individual pulsar remain intact. But all phase coherence between pulsars is eliminated. The second technique involves scrambling (not merely interchanging amongst other pulsars) the pulsar positions used to construct the \textit{Hellings and Downs} ORF template in our search models. This is designed to be orthogonal to the true signal's ORF so that correlations between pulsars are destroyed. Both of these techniques operate on the true, measured PTA dataset rather than on noise simulations. This incorporates all idiosyncrasies of the true dataset into assessing the detection statistic significance, instead of being biased by our (possibly incomplete) noise model assumptions.

We tested our techniques against several different types of PTA datasets, including $(1)$ an idealized dataset (with a large number of evenly sampled pulsars, high timing precision, and no intrinsic red noise); $(2)$ a more realistic dataset (realistic cadence, timing precision, and red noise levels); and $(3)$$+$$(4)$ datasets for which our noise model is incomplete (i.e.\ includes either intrinsic or spatially-correlated noise processes which we do not explicitly model). By performing several hundred phase-shifting and sky-scrambling analyses, we constructed a distribution of the Bayes factor for spatial correlations under the null hypothesis, which allows us to quote the $p$-value of the true Bayes factor. Quoting $p$-values when our detection statistic is Bayesian may seem like an ill-conceived mixture of two distinct inference philosophies, but it is merely trying to answer the question of what our recovered Bayes factor actually means, and how often noise alone could produce it. Our techniques can also easily be used with frequentist detection statistics, in which case there is no conflict of philosophies.

For the idealized and realistic datasets, we found that the distribution of Bayes factors produced by phase-shifting and sky-scrambling compared well with that of analyzing many noise simulations. We took a further step in showing that our two techniques are actually superior to noise simulations in the case where we have a poor or incomplete noise model. Noise simulations will include only the processes of our incomplete noise model, and so will provide a null hypothesis distribution which exaggerates the significance of the true Bayes factor. Phase-shifting operates on the real PTA dataset and provides a more conservative estimate of detection significance, while sky-scrambling is found to give a significance somewhere between the noise-simulation and phase-shifting results. For this reason, and the fact that there are limitations on the number of independent sky-scrambles we can make, the more general and reliable approach appears to be phase-shifting.

These techniques can be readily deployed on all existing and future PTA datasets, and should become standard tools for contextualizing the GWB Bayes factors that we report in future PTA analyses. They already exist as modeling options within the Bayesian pulsar-timing analysis package, NX01 \citep{nx01}. Understanding the significance of our quoted detection statistics is of vital importance as PTAs move closer to the first detection of nanohertz gravitational waves. Growing the GWB detection significance needs strategies that are designed to resolve the spatial correlations between pulsars. This requires many well-timed pulsars with long observational baselines, and which are widely separated across the sky. Without broad sky coverage, a PTA will not be able to distinguish between a GWB, clock errors, or potentially other spatially-correlated processes. The International Pulsar Timing Array plays a vital role in this effort by pooling observations and pulsars from the individual PTAs, and coordinating a unified international GW search program.

\begin{acknowledgments}
We thank the anonymous referee for many useful suggestions that improved the quality of this paper. We also thank Justin Ellis, Michele Vallisneri, Rutger van Haasteren, Joseph Lazio, Neil Cornish, Laura Sampson, Paul Demorest, G.~J.~Babu, as well as the entire NANOGrav and EPTA detection working groups for many useful suggestions and fruitful discussions. SRT was supported by appointment to the NASA Postdoctoral Program at the Jet Propulsion Laboratory, administered by Oak Ridge Associated Universities and the Universities Space Research Association through a contract with NASA. AS is supported by a University Research Fellowship of the Royal Society. The authors acknowledge the support of colleagues in the EPTA. This work was supported in part by National Science Foundation Grant No. PHYS-1066293 and by the hospitality of the Aspen Center for Physics. A majority of the computational work was performed on the Nemo cluster at UWM supported by NSF grant No. 0923409. The research was partially carried out at the Jet Propulsion Laboratory, California Institute of Technology, under a contract with the National Aeronautics and Space Administration. \copyright\,~2016. All rights reserved.
\end{acknowledgments}

\bibliography{apjjabb,noiseChar}

\end{document}